\documentclass[twocolumn,a4paper]{article}

 \usepackage{pstricks}
 \usepackage{graphicx}
 \usepackage{dblfloatfix}

 \usepackage{xfrac}

 \newcommand{\Aai}{\textnormal{\AA}^{-1}}

\begin{document}
%===============================================================================
  \title{%
   High intensity specular reflectometry \\---\\ first experiments
        }
  \author{%
   J.\ Stahn$^a$
   U.\ Filges$^b$
   T.\ Panzner$^b$
  }
  \date{% 
   $^a$Laboratory for Neutron Scattering,
   and
   $^b$Laboratory for Developments and Methods,  \\[0.5ex]
   Paul Scherrer Institut, 5210 Villigen PSI, Switzerland
  } 
  \maketitle
  {\small%
   \noindent
   corresponding author: Jochen Stahn,\\
   e-mail: jochen.stahn@psi.ch, \\
   tel: +41\,56\,310\,2518, 
   fax: +41\,56\,310\,2939
  }
%-------------------------------------------------------------------------------
\begin{abstract}
  \vspace{-1ex}
 Selene is the attempt to implement a new scheme for high-intensity
 specular reflectometry. Instead of a highly collimated beam one uses
 a convergent beam covering a large angular range. The angular resolution
 is then performed by a position-sensitive detector. Off-specular scattering
 in this set-up leads to some background, but for screening of wide
 parameter ranges (e.g. temperature, electric and magnetic fields)
 the intensity gain of at least one order of magnitude is essential.
 If necessary, the high precession measurements (even with off-specular
 components) then are performed with the conventional set-up.
 The heart of this new set-up is an elliptically focusing guide element
 of 2\,m length. Though this guide is optimised for the use on the TOF
 reflectometer Amor at SINQ, it can be used as stand-alone device
 to check the possible application also for other neutron scattering
 techniques.
 The first measurements on AMOR confirmed the general concept and the
 various operation modes. A draw-back occurred due to problems with the
 internal alignment of the guide. Nevertheless in the TOF mode a gain factor
 of 10 was reached, and a factor 25 can be expected for an improved guide.

 keywords:
 neutron, reflectometry, elliptic guide, focusing
\end{abstract}

%-------------------------------------------------------------------------------
\section{the ideal}

 In a first step we developed an \textsl{ideal} neutron reflectometer,
 based on the convergent beam concept by Ott \cite{Ott200823}
 to gain intensity for specular reflectometry on small samples.
 The idea to use half an elliptic guide for this purpose
 has already been published.\cite{Ott200823,Stahn2010}
 But the idea of using elliptic guides is much older and has
 been tested and realised several times in the past.
 \cite{Muehlbauer200877,mue06a}

 Besides this approach, there are other concepts on the market 
 \cite{Ott2008401,Ott2009EPJ} and under investigation \cite{cub06}
 to skip the off-specular signal to decrease the 
 measurement times for specular reflectometry.

 %..............................................................................
 \subsection{principles}

  The design of the reflectometer study \textsl{selene} is based on the 
  following principles:
  \begin{itemize}
   \item
    \textsl{The beam characteristics are defined at the sample}. 
    This means that the expected sample size and the acceptable or
    intended divergence and wavelength range are identified, and that
    all components are optimised to operate within these ranges.
    A consequence is e.g.\ that the coating of a guide might
    be optimised for high reflectance instead of high reflecting angles.
   \item
    \textsl{Useless neutrons have to be avoided},
    since they cause radiation and background problems. 
    This is essentially true for the avoidable illumination of the sample 
    environment, caused by a divergent beam. The smaller the
    sample the more one profits of using a focused beam.
   \item
    \textsl{Any beam filtering / shaping should be performed as 
    early as possible}. Again this leads to lower background, 
    reduces radiation problems, and the shielding design can be
    simplified.
    E.g.\ the fraction of neutrons finally arriving at the sample 
    at a reflectometry set-up is of the order $10^{-4}$ compared
    to the intensity leaving a conventional straight guide. 
    In this case the neutrons not used are absorbed or scattered
    within a few meters from the sample (and detector).
   \item
    \textsl{Optimisation vs. universality}: 
    Every degree of freedom to have a more flexible instrument
    leads to some compromise. So often a multi-task instrument 
    becomes quite complicated and does not allow for state of the
    art measurements. It might be cheaper and more efficient to
    build several \textsl{simple} but highly specialised instruments.
  \end{itemize}

  In the present case the instrument is optimised for
  \textsl{specular reflectometry on small samples}. Where small
  means surface area in the range of some $\textnormal{mm}^2$. For these samples 
  it is very demanding and time-consuming to measure off-specular
  reflectivity and in most cases one is content to obtain specular
  reflectivity curves for a few external parameters, only. A screening
  of larger temperature or magnetic or electric field ranges is
  very time consuming.  

    \begin{figure}[b]
          \begin{picture}(80,26)
      \put(-15,-5){
        \psset{unit=0.5mm,linestyle=none}
        \psset{fillstyle=solid}
         \definecolor{gb1}{rgb}{1.0,0,0.0}  \definecolor{ge1}{rgb}{1.5,0.5,0.5}
         \definecolor{gb2}{rgb}{0.9,0,0.1}  \definecolor{ge2}{rgb}{1.4,0.5,0.6}
         \definecolor{gb3}{rgb}{0.8,0,0.2}  \definecolor{ge3}{rgb}{1.3,0.5,0.7}
         \definecolor{gb4}{rgb}{0.7,0,0.3}  \definecolor{ge4}{rgb}{1.2,0.5,0.8}
         \definecolor{gb5}{rgb}{0.6,0,0.4}  \definecolor{ge5}{rgb}{1.1,0.5,0.9}
         \definecolor{gb6}{rgb}{0.5,0,0.5}  \definecolor{ge6}{rgb}{1.0,0.5,1.0}
         \definecolor{gb7}{rgb}{0.4,0,0.6}  \definecolor{ge7}{rgb}{0.9,0.5,1.1}
         \definecolor{gb8}{rgb}{0.3,0,0.7}  \definecolor{ge8}{rgb}{0.8,0.5,1.2}
         \definecolor{gb9}{rgb}{0.2,0,0.8}  \definecolor{ge9}{rgb}{0.7,0.5,1.3}
         \definecolor{gb10}{rgb}{0.1,0,0.9} \definecolor{ge10}{rgb}{0.6,0.5,1.4}
         \definecolor{gb11}{rgb}{0.0,0,1.0} \definecolor{ge11}{rgb}{0.5,0.5,1.5}
        \psline[fillcolor=gb1](100,30)(50,10)(100,33)
        \psline[fillcolor=gb2](100,33)(50,10)(100,36)
        \psline[fillcolor=gb3](100,36)(50,10)(100,39)
        \psline[fillcolor=gb4](100,39)(50,10)(100,42)
        \psline[fillcolor=gb5](100,42)(50,10)(100,45)
        \psline[fillcolor=gb6](100,45)(50,10)(100,48)
        \psline[fillcolor=gb7](100,48)(50,10)(100,51)
        \psline[fillcolor=gb8](100,51)(50,10)(100,54)
        \psline[fillcolor=gb9](100,54)(50,10)(100,57)
        \psline[fillcolor=gb10](100,57)(50,10)(100,60)
        \psline[fillcolor=gb11](100,60)(50,10)(100,63)
        \psline[fillcolor=gb1](0,30)(50,10)(0,33)
        \psline[fillcolor=gb2](0,33)(50,10)(0,36)
        \psline[fillcolor=gb3](0,36)(50,10)(0,39)
        \psline[fillcolor=gb4](0,39)(50,10)(0,42)
        \psline[fillcolor=gb5](0,42)(50,10)(0,45)
        \psline[fillcolor=gb6](0,45)(50,10)(0,48)
        \psline[fillcolor=gb7](0,48)(50,10)(0,51)
        \psline[fillcolor=gb8](0,51)(50,10)(0,54)
        \psline[fillcolor=gb9](0,54)(50,10)(0,57)
        \psline[fillcolor=gb10](0,57)(50,10)(0,60)
        \psline[fillcolor=gb11](0,60)(50,10)(0,63)
        \psline[fillstyle=solid,fillcolor=green](30,10)(70,10)(70,5)(30,5)
        \psset{fillstyle=none,linestyle=solid}
        \psarc[linecolor=white,linewidth=20](50,10){63}{20}{50}
        \psarc[linecolor=white,linewidth=20](50,10){63}{130}{160}
        \psarc[linecolor=black,linewidth=1]{->}(50,10){60}{130}{160}
        \psarc[linecolor=black,linewidth=1]{<-}(50,10){75}{130}{160}
        \psarc[linecolor=black,linewidth=1]{->}(50,10){60}{20}{50}
        \put(-11,25){\small$\alpha_i^\textnormal{s}$}
        \put(-3,22){\small$\lambda$}
        \put(47,26){\small$\displaystyle
                    q_z \propto 
                     \frac{\sin\alpha_i^\textnormal{s} }
                          { \lambda(\alpha_i^\textnormal{s})}$}
      }
    \end{picture}
     \caption{\label{f_refocus}
      Principle of REFocus in the dispersive mode: an incoming convergent beam with a
      $\lambda(\alpha_i)$ encoding is specularly reflected off the sample
      (\textsl{green}). Since $\alpha_i$ and $\lambda$ increase in the opposite
      direction, a large $q_z$ range is covered simultaneously. In the case where
      $\alpha_i$ and $\lambda$ increase in the same direction one has still 
      dispersion, but at a much reduced range.
     }
    \end{figure}
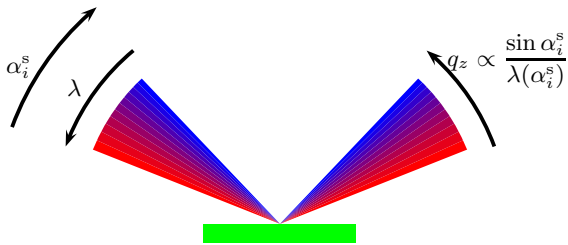 
 
  To shorten the measurement time for specular reflectivity curves, at least
  for screening purposes, F.\ Ott suggested to use a beam converging to the
  sample and covering a large range of incidence angles.\cite{Ott200823,Ott2009EPJ} 
  This concept, called REFocus, can be assessed as an angle dispersive reflectometer
  set-up, but with a strongly converging beam (the sample is
  in the focal point) with a relation between the angle of incidence 
  on the sample $\alpha_i^\textnormal{s}$
  and the wavelength $\lambda = \lambda(\alpha_i^\textnormal{s}$). 
  The focusing and $\lambda$ selection is performed by
  one branch of an elliptically shaped guide element (acting as a lense)
  which is coated with a monochromatising multilayer (ML).
  The period of the ML might be graded along the guide.
  For specular conditions on the sample one has the exit angle 
  $\alpha_f^\textnormal{s} = \alpha_i^\textnormal{s}$
  and thus the normal neutron momentum transfer
  $q_z = 4\pi \sin\alpha_f^\textnormal{s} / \lambda(\alpha_f^\textnormal{s})$.   
  Figure \ref{f_refocus} illustrates the situation on the sample.
  The function $\lambda(\alpha_f^\textnormal{s})$ is given by the shape of the
  guide element and its coating.
  The measured quantity $I(\alpha_f^\textnormal{s})$ can be converted into
  reflectivity $R(q_z)$. 
  Since a broad $\alpha_i^\textnormal{s}$-range is active simultaneously, 
  the $\alpha_f^\textnormal{s}$- and thus $q_z$-resolution is given by the 
  spacial resolution and distance to the sample of a position-sensitive 
  detector (PSD).  
  Assuming a homogeneous intensity distribution over the 
  $\alpha_i^\textnormal{s}$ range, the intensity gain relative to a 
  conventional set-up with small $\Delta \alpha_i^\textnormal{s}$ is 
  proportional to $\Delta \alpha_{i,\mathrm{REFOCUS}}^\textnormal{s} 
  / \Delta \alpha_{i,\mathrm{conventional}}^\textnormal{s}$, which is $>10$.

  Depending on the orientation of the sample towards the incoming beam
  and the gradient of the ML-coating one has more or less dispersive geometries, 
  depending on the needs of the experiment.
  F.\ Ott expects to cover a dynamic range of about $10^5$.
  The limitation is the off-specular scattering from the sample 
  leading to a non-flat background.

  REFocus fulfils the requirement for a convergent beam and thus
  a rather small illumination of the sample environment.
  But the monochromatisation is performed all along the guide and thus
  one has still a white beam close to the end of the guide.

 %..............................................................................
 \subsection{selene concept}

  Based on REFocus we developed the selene concept to allow only those
  neutrons to enter the guide which are actually used at the sample.
  Since the ellipse ideally maps the pre-image (the virtual source) 
  at one focal point
  to the adjoint one, this means that the beam should have the required
  properties already at the first focal point.
  This can be realised \dots \\
  \begin{enumerate}
  \item[(I)]
        \dots with a multilayer monochromator, where for each incident angle one has
        a different, but defined energy ($\lambda(\alpha_i)$ encoding):
        A ML-monochromator (reflecting at $q_z^m$) or a ML-bandpass is located
        directly in front of, or even at the first focal point. 
        In combination with a small slit or knife-blade diaphragm it defines the 
        virtual source to be mapped to the sample.
        The incoming divergent white beam is reflected
        from that monochromator if $4\pi \sin \alpha_i^m / \lambda \in \{q_z^m\}$.
        This means that for a given incoming-angle on the monochromator
        $\alpha_i^m$ a wavelength band according to the coating of the
        monochromator is reflected into $\alpha_f^m$ --- which by the
        elliptic guide is transformed into an $\alpha_i^s$ at the sample.
        In the following this operation mode will be referred to as 
        \textsl{monochromatic}.
        \\
  \item[(II)]
        \dots in time-of-flight (TOF) mode, where at a given time all neutrons
        have the same energy:
        Here the wavelength is encoded in $\lambda = \lambda(\mathrm{TOF})$.
        As a consequence, optical errors of the monochromator
        or the elliptic guide (waviness, misalignment) do not influence the 
        energy resolution, but just reduce the intensity on the sample.
        \\
  \item[(III)]
        \dots with a crystal monochromator, with a fixed and constant energy:
        This is conceptional the simplest approach, but in the
        present case the most difficult to realise. The geometry of the
        instrument would require crystals with a large mosaicity, and
        a double monochromator set-up. This is because one wants to avoid
        moving the elliptic guide when changing the wavelength.
        This set-up will be investigated in the future.
  \end{enumerate}

  Options I and II were tested and will be discussed in more detail.
  The actual $\lambda$-range is optimised due to the $I(\lambda)$
  distribution of the source. 
  In the sample plane the resolution
  conditions are relaxed and focusing is also favourable. But here 
  the sample is no longer \textsl{point-like}, so it is more
  difficult to avoid over-illumination.

  For small samples the virtual source does not have to be much larger than the
  projected height of the sample, i.e. it is in the sub-mm range
  --- independent of the size
  of the complete guide. In the extreme case this means that a 
  guide for a reflectometer needs an incoming aperture close to
  the source of some mm only, instead of the 30 to 50\,mm used in conventional
  guides.
  
  If off-specular measurements are required, or if the sample
  does not allow for using a large $\Delta\alpha_f$, it is possible to
  return to a configuration very close to the conventional angle-dispersive
  set-up. By reducing the divergence (and thus the $\lambda$-range)
  by a slit one gets a close-to-monochromatic beam and a high angular
  resolution. The intensity will be still slightly higher than
  in the conventional geometry due to less reflections in the guide.
  And the beam on the sample is still convergent and does not over-illuminate
  the sample more than necessary.

 %..............................................................................
 \subsection{aberration and correction}
 
    \begin{figure}[t]
        \begin{picture}(80,45)
    \psset{unit=1mm}
    \put(39,28){% coma
      \psset{linestyle=solid,fillstyle=none,linewidth=0.2,labelsep=1mm}%
      \psline[linecolor=black](-35,0)(-20,7.5)(35,0)
      \psline[linecolor=blue](-35,-5)(-20,7.5)(35,-10)
      \psline[linecolor=black](-35,0)(20,7.5)(35,0)
      \psline[linecolor=green](-35,-5)(20,7.5)(35,-2.5)
      \psline[linecolor=black,linewidth=0.5](-35,0)(-35,-5)
      \psline[linecolor=blue,linewidth=0.5](35,-10)(35,0)
      \psline[linecolor=green,linewidth=0.6](35,-2.5)(35,0)
      \psellipticarc[showpoints=false,linecolor=red,linewidth=0.4]{-}(0,0)(35,10){0}{180} 
      \uput[180](-35,-5){\small P}
      \uput[0](35,-2.5){\small \color{green}P'}
      \uput[0](35,-10){\small \color{blue}P'}
    }
    \put(39,2){% (de)focusing
      \psset{linestyle=none,fillstyle=solid}%
      \psline[fillcolor=blue](-35,0)(-22,7.5)(-18,8)(35,0)(-22,7.5)(-18,8)(-35,0)
      \psline[fillcolor=green](-35,0)(15,8.5)(25,7)(35,0)(15,8.5)(25,7)(-35,0)
      \psellipticarc[showpoints=false,fillcolor=green]{-}(0,0)(35,10){140}{160} 
      \psset{linestyle=solid,fillstyle=none,linewidth=0.4}%
      \psellipticarc[showpoints=false,linecolor=red]{-}(0,0)(35,10){0}{180} 
    }
   \end{picture}
     \caption{\label{f_coma}
      Sketches to illustrate coma aberration. 
      \textsl{Top:}
      A beam from an off-axis point P of the pre-image is reflected off the reflector
      with an angular offset with respect to the ideal beam (from the focal point).
      This angular offset varies along the reflector: in the early part (\textsl{blue})
      it is large, in the late part (\textsl{green}) it is small.
      As a result the reflected beams
      hit the second focal plane at various points P', the image is blurred.
      \textsl{Bottom}:
      On the other hand side beams from the first focal point emitted in a defined
      solid angle are reflected by a short / long part of the reflector in its 
      early / late (\textsl{blue} / \textsl{green}) part. 
      The result is that the beam is focused / de-focused.
      The combination of both effects leads to a distortion of the phase space,
      where its density stays constant.
     }
    \end{figure}
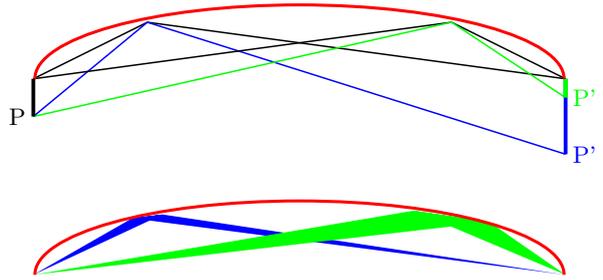
  
  Focusing guides show aberration effects. In case of an elliptic guide
  it is coma aberration, i.e. an off-axis point of the pre-image is projected 
  onto a line in the image plane. 
  The amount and direction of the distortion depends on where along the
  ellipse the reflection occurs.
  Figure \ref{f_coma} displays this effect.

  The consequence is that (I) the size of the spot in the image plane
  is larger than the pre-image (the slit), defined by reflection
  in the early part of the guide; and (II) the size of the spot with
  a constant intensity for all incoming angles is smaller than the
  pre-image. This is caused by the focusing effect of reflections close
  to the end of the guide.
  For typical elliptic guides and sample sizes it turned out that 
  due to the coma aberration the virtual source size has to be about 3 times
  the sample size.  
  Then the complete sample is illuminated with the
  same divergence. The beam spot at the sample is then about
  10 times the sample size, but the intensity drops fast outside
  the inner homogeneous region.
  One can estimate that about 30\% of the beam hits the sample,
  the rest leads to illumination of the sample environment.
 
  While this is already much less than compared to the usage of a
  divergent beam, it still leads to background problems. It is possible
  to correct for the coma aberration by dividing the guide into 
  two identical elliptic parts which have one focal point in common.
  The coma aberration of the first guide leads to a blurred 
  intermediate image in the joining focal plane. Due to the reversibility
  of the optical paths, this image is converted back to a sharp image 
  at the sample position. 
  Neglecting the limited reflectivity
  of the guides, one gets a beam at the sample position with almost the
  size given by a slit at the first focal point and the divergence
  defined by the acceptance of the elliptic guide.

  P.\ B\"oni suggested to use subsequent elliptic guides, joining a 
  focal point to allow for small beam-shaping elements like
  choppers and RF-coils.\cite{Boeni2007NIM} In the cases with an 
  even number of ellipses, coma aberration is no problem. 

 %..............................................................................
 \subsection{castle in the air}

  Disregarding real constraints like the finite width of radiation shielding
  and the non-perfect SM coating one can put together the items
  mentioned above to build an ideal reflectometer for small samples.

  In detail, along the beam path:
  The early filtering means that a monochromator or chopper is placed
  directly behind the neutron source (in a way that a large divergence
  is collected). A very small first slit (opening exactly the projected
  height of the sample) forms the $\mathrm{1^{st}}$ focal point of first elliptic 
  reflector. In the $\mathrm{2^{nd}}$ focal point the beam can be manipulated further,
  e.g. it can be polarised or a $\mathrm{2^{nd}}$ chopper can be installed. Here
  the image is distorted by coma aberration, and (if relevant) the
  $\lambda(\alpha_i)$ encoding is not linear. 
  A $\mathrm{2^{nd}}$ elliptic reflector with the same 
  parameters as the first follows. Both reflectors share the 
  $\mathrm{2^{nd}}$ focal point.
  The sample will be positioned in the $\mathrm{3^{rd}}$ focal point. 
  Here aberration effects are almost cancelled and a quasi-linear
  $\lambda(\alpha_i)$ encoding is restored.
  A high-resolution ($<0.5\,\textnormal{mm}$) 
  PSD is positioned about 500\,mm behind the sample.

%-------------------------------------------------------------------------------
\section{the reality}

 The limited amount of resources and the absence of access to 
 a free and freely configurable beam-port at a could source tell, that it
 is not realistic to test the instrument concept mentioned above by just
 building it. 
 Instead we decided to build a down-scaled simplified version to be tested
 on the TOF reflectometer Amor at SINQ, PSI. Amor consists of an
 optical bench on which all components can be positioned with rather high
 flexibility. In short, the "cold source" is replaced by the end of
 the neutron guide; the ML-monochromator is positioned as close as
 possible to the end of the guide; a 2\,m long elliptically shaped
 deflector is positioned in between monochromator and sample; and the
 detection is performed by the PSD.

 This set-up restricts the available divergence to $\approx 1.2^\circ$
 in the monochromatic mode and to $\approx 1.6^\circ$ for the TOF mode,
 respectively.
 Coma aberration is not corrected for. Nevertheless, it is
 possible to measure the gain factors for specular reflectivity;
 the fall-back option to the off-specular angle dispersive mode; and
 the problems caused by diffuse scattering from the monochromator.

 %..............................................................................
 \subsection{experimental environment: Amor}

  Amor is a neutron reflectometer which allows for a wide range of 
  set-ups.\cite{gup04}
  The scattering geometry is vertical so that liquid surfaces are accessible.
  Most components are positioned on an optical bench which allows to play
  with the resolution, or to test exotic set-ups like the prism approach
  by R.\ Cubitt \cite{cub06} or the selene concept. 
  In general Amor is operated in TOF mode (realised by a double
  chopper), but it is also possible to run it with a monochromator.

  For the tests presented here, the
  elliptic guide is positioned on the sample table (which allows for all
  necessary degrees of freedom) and the sample is on the analyser stage.
  The only restriction resulting from this shift is that the sample environment
  has to be small and light.
  The ML-monochromator is placed on the polariser stage and can be driven out 
  of the beam to allow for using selene in TOF mode.
  
  In the following sections the components are discussed in detail.
  The following conventions have been used:
  The laboratory coordinate system is 
  right handed with $x$-axis horizontally along the beam, 
  and $z$-axis vertical. Local coordinate systems follow this
  convention as much as possible. E.g.\ at the sample $z$ is normal
  to the surface and $x$ in the surface; both tilted to the laboratory
  system by only a few degrees. $y$ is unchanged.
  For defining the various angles (relative to the lab system, the
  local system and eventually including deviations) a reference
  beam is introduced. It leaves the neutron guide at mid-height
  horizontally ($x$-direction). After that it is assumed to be specularly
  reflected in the $xz$-plane at all optical surfaces, i.e.\
  by the double monochromator by $2\theta^\textnormal{m}$ 
  and $-2\theta^\textnormal{m}$;
  by the centre of the elliptic guide by $2\theta^\textnormal{e}$; 
  and on the sample by $2\theta^\textnormal{s}$.
  $\theta$ denotes the angle between the (reflecting) surface and the 
  incoming reference beam, whereas $\alpha$ is the angle of some beam 
  relative to the surface.
  The superscripts mean \textsl{m}\/onochromator, 
  \textsl{e}\/lliptic guide, \textsl{s}\/ample, and 
  \textsl{d}\/etector.

 %..............................................................................
 \subsection{chopper}

  The chopper is positioned in a housing, some cm behind the end
  of the neutron guide. It consists of 2 discs, 490\,mm apart, each with 2
  openings of $13.6^\circ$. In general it is operated in a way to
  give $\Delta\lambda / \lambda = \textnormal{const.}$\cite{Well1992}
  
  The chopper housing limits the maximum incoming divergence 
  because it leads to a minimum distance between the end of the guide
  to the first diaphragm of 1500\,mm. And to the first place where 
  a monochromator can be installed (on the frame overlap filter stage) 
  it is some 1800\,mm. 

  For the tests we used a pulse
  frequency of $23.\bar{3}\,\textnormal{Hz}$ (corresponding to 700\,rpm).

 %..............................................................................
 \subsection{multilayer monochromator}\label{s_mono}

    \begin{figure*}[t!]
     \center{\includegraphics[width=158mm]{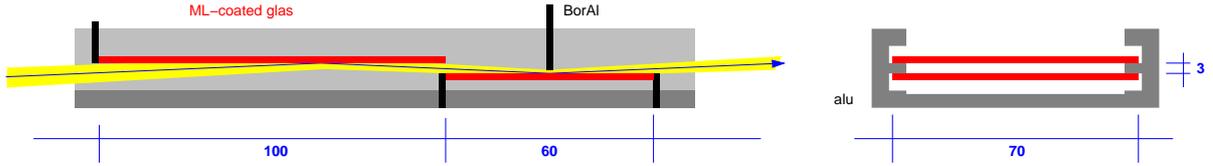}}
      \caption{\label{f_ml_momo_sketch}
       Cuts along (\textsl{left}) and normal (\textsl{right})
       to the beam path through the double 
       ML-monochromator. \textsl{Red} means glass substrate, \textsl{black} is the
       BorAl absorber and \textsl{yellow} the beam (coming from the left side).
      }
    \end{figure*}

    \begin{figure}[t]
     \input{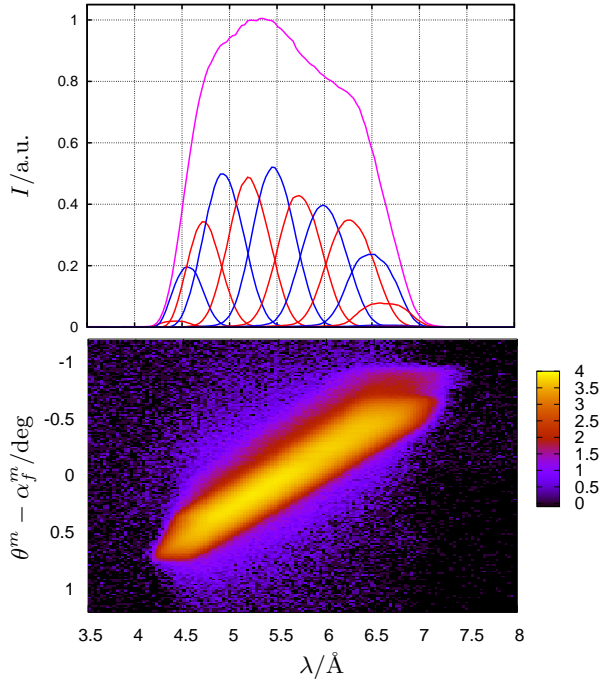}
     \caption{\label{f_mono_div}
      \textsl{Bottom}: Intensity map
      for $\theta^m = 3^\circ$ and
      $\Delta\alpha_i^m \approx 1.4^\circ$ on a 
      logarithmic scale. 
      $\log_{10}I(\lambda,\alpha_f^m)$ 
      \textsl{Top}: Various cuts through that map along $\lambda$ 
      (\textsl{alternating red and blue}), 
      and the total intensity as function of $\lambda$, integrated over 
      $\alpha_f^m$ (\textsl{magenta}).
     }
    \end{figure}

   The monochromator is designed for a fixed geometry. 
   The coating is a Ni/Ti bandpass with a plateau in the range 
   $q_z^m \in [ 0.1\,\Aai, 0.11\,\Aai ]$. This corresponds to
   $m \in [4.5, 5.0]$. 

   Two glass substrates coated with the mentioned ML are arranged
   face to face with a 3\,mm gap to form a double monochromator.
   The glasses are shifted so that they just do not overlap
   (see figure \ref{f_ml_momo_sketch}).
   A knife blade atop the middle of the $\mathrm{2^{nd}}$ mirror
   is used to define the width of the beam in the $\mathrm{1^{st}}$
   focal point of the ellipse.

   The double monochromator was positioned on the
   polariser stage, 2744\,mm after the end of the guide. 
   Thus the maximum divergence to be expected there is 
   $\Delta\alpha_i^m = 1.05^\circ$.
   This is the set-up shown in figure \ref{f_amor_selene_mono}.

   For this incoming divergence one gets a reflected beam with
   $\lambda \in [4.2\,\textnormal{\AA}, 7\,\textnormal{\AA}]$, and
   $\Delta\lambda / \lambda \approx 9\%$, only for  
   $\lambda < 4.4\,\textnormal{\AA}$ the resolution gets better
   due to cut off effects, as can be seen in figure \ref{f_mono_div}.
   This resolution still contains the $\Delta\lambda / \lambda \approx
   4\%$ given by the chopper settings.
 
   The intensity distribution for various $\alpha_f^m$ shown 
   in figure \ref{f_mono_div}, top, originates from the 
   $I(\lambda)$ of the incident beam and for smaller
   $\lambda$ (i.e.\ smaller $\alpha_i^m$) also from the
   limited length of the monochromator.

   Using this set-up with a monochromating ML offers the opportunity
   to switch easily to polarised neutrons by just replacing the 
   ML by a polarising coating.\cite{Stahn2004243}

 %..............................................................................
 \subsection{elliptic guide}

    \begin{figure*}[t!]
         \begin{picture}(100,50)
     \unitlength 0.666mm
     \psset{unit=0.666mm, xunit=0.020mm, yunit=0.20mm}
     \put(135,45){
       % boxes & shadows
       \psset{linestyle=none,fillstyle=solid}%
       \psline[fillcolor=yellow](-4000,0)(-3000,43.3)(-2700,48)(-2000,52)%
               (-1300,48)(-1000,43.3)%
       \psline[fillcolor=yellow](-3000,43.3)(-1000,43.3)(0,0)%
       \psline[fillcolor=gray](-3000,0)(-2004,0)(-2004,15)(-3000,15)% 
       \psline[fillcolor=gray](-1000,0)(-1996,0)(-1996,15)(-1000,15)% 
       % absorber
       \psset{linestyle=solid,fillstyle=none,linecolor=black,linewidth=0.06}%
       \psline(-2000,30)(-2000,-10)% knife selene
       % the grid
       \psset{linestyle=solid,fillstyle=none,linecolor=blue,linewidth=0.01}%
       \psline{->}(-4200,0)(200,0)%
        \uput[-35](200,0){\small\color{blue}$x / \textnormal{mm}$}%
       \psline[linestyle=dotted](-4000,55)(-4000,-2)%
        \uput[-90](-4000,0){\small\color{blue}-4000}%
       \psline[linestyle=dotted](-3000,55)(-3000,-2)%
        \uput[-90](-3000,0){\small\color{blue}-3000}%
       \psline[linestyle=dotted](-1000,55)(-1000,-2)%
        \uput[-90](-1000,0){\small\color{blue}-1000}%
       \psline{<-}(  0, 60)(  0,-2)%
        \uput[-135](    0,0){\small\color{blue}0}%
        \uput[35](0,60){\small\color{blue}$z / \textnormal{mm}$}%
       \psline[linestyle=dotted]{-}(-2000,50)(20,50)%
        \uput[0](20,50){\small\color{blue}$50$}%
       \psline[linestyle=dotted]{-}(-1000,15)(20,15)%
        \uput[0](20,15){\small\color{blue}$15$}%
       % guides
       \psellipticarc[linewidth=0.1,showpoints=false,linecolor=blue,linestyle=dotted]%
           {-}(-2000,0)(2000,50){0}{180}%
       \psarc[linewidth=0.1,showpoints=false,linecolor=blue]{->}(-4000,0){20}{0}{8}%
        \uput[90](-3350,30){\small\color{blue}$2.479^\circ$}%
       \psarc[linewidth=0.1,showpoints=false,linecolor=blue]{->}(-4000,0){25}{0}{24}%
        \uput[90](-3450,5){\small\color{blue}$0.827^\circ$}%
       \psset{linecolor=red}%
       \psellipticarc[linewidth=1.5,showpoints=false]{-}(-2000,0)(2005,55){60}{120}%
       % labels
        \uput[90](-2000,  55){\small\color{red}elliptic shape, coating: Ni/Ti $m=5$}%
        \uput[90](    0,  70){\small\color{black}sample}%
        \uput[90](-4000,  70){\small\color{black}monochromator}%
     } %off-set
     %-----------
     \put(135,13){
       % boxes & shadows
       \psset{linestyle=none,fillstyle=solid}%
       \psline[fillcolor=yellow](-4120,0)(-3000,40.3)(-2700,44)(-2000,48)%
               (-1300,44)(-1000,38.6)(0,0)(-1000,-38.6)(-1300,-44)%
               (-2000,-48)(-2700,-44)(-3000,-40.3)(-4120,0)
       % absorber
       \psset{linestyle=dashed,fillstyle=none,linecolor=black,linewidth=0.06}%
       \psline(-2000,46)(-2000,-46)% knife selene
       % the grid
       \psset{linestyle=solid,fillstyle=none,linecolor=blue,linewidth=0.01}%
       \psline{->}(-4200,0)(200,0)%
        \uput[-35](200,0){\small\color{blue}$x / \textnormal{mm}$}%
       \psline[linestyle=dotted](-4000,-50)(-4000,90)%
       \psline[linestyle=solid](-4000,-5)(-4000,5)%
       \psline[linestyle=dotted](-3000,-50)(-3000,90)%
       \psline[linestyle=dotted](-1000,-50)(-1000,90)%
        \uput[-135](    0,0){\small\color{blue}0}%
       \psline{<-}(  0, 60)(  0,-50)%
        \uput[35](0,60){\small\color{blue}$y / \textnormal{mm}$}%
       \psline[linestyle=dotted]{-}(-2000,45)(20,45)%
        \uput[0](20,45){\small\color{blue}$45$}%
       \psline[linestyle=dotted]{-}(-2000,-45)(20,-45)%
        \uput[0](20,-45){\small\color{blue}$-45$}%
       \psarc[linewidth=0.1,showpoints=false,linecolor=blue]{<-}(0,0){20}{160}{180}%
        \uput[90](-450,20){\small\color{blue}$2.21^\circ$}%
       % guides
       \psellipticarc[linewidth=0.1,showpoints=false,linecolor=blue,linestyle=dotted]%
           {-}(-2060,0)(2060,45){0}{360}%
       \psset{linecolor=red}%
       \psellipticarc[linewidth=1.5,showpoints=false]{-}(-2060,0)(2060,50){58.5}{117.5}%
       \psellipticarc[linewidth=1.5,showpoints=false]{-}(-2060,0)(2060,50){242.5}{301.5}%
       % labels
        \uput[90](-2000,  50){\small\color{red}elliptic shape, coating: Ni/Ti $m=3.5$}%
     } %off-set
    \end{picture}
     \hfill\parbox[b]{50mm}{
     \caption{\label{f_guide}
      Sketch of the elliptic guide with dimensions. The $y$ and $z$
      scales are stretched by a factor 10 relative to the $x$ scale.
      \textsl{Top}: cut in the scattering plane ($x$-$z$-pane),
      \textsl{bottom}: cut in the sample plane ($x$-$y$-pane).
     }
     }
    \end{figure*}
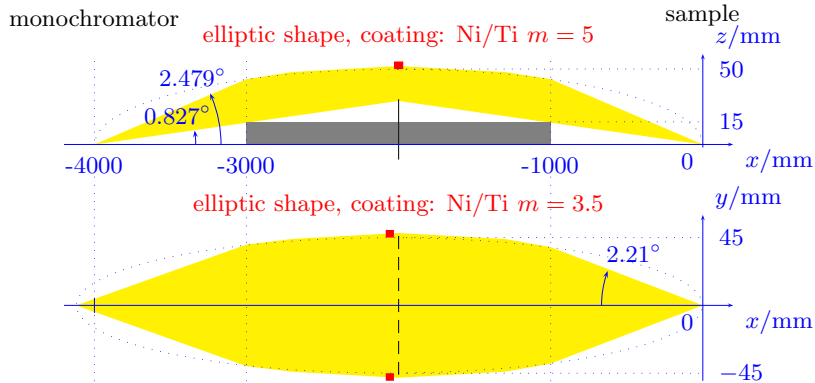

  The guide has two functions: it has to focus the beam in 
  the sample ($xy$) plane
  to the sample without taking care for the divergence; and it has to
  map the pre-image to the sample in the scattering ($xz$) plane, keeping
  a precise $\alpha_i$/$\lambda$ correlation. 

  The first aim is accomplished by the elliptic shape of the side-walls
  of the guide, where the second focal point is on the sample, while
  the first focal point is before the pre-image. The reason for this is
  that in the $y$-direction the sample is not really small with respect to
  the guide dimensions. Typical sample widths are 2 to 10\,mm. 
  Taking aberration into account this would mean a pre-image width of up to
  30\,mm --- which is more than half of the length of short axis of the
  ellipse. 
  The half axes parameters are $a_{xy} = 2025\,\textnormal{mm}$ and
  $b_{xy} = 45\,\textnormal{mm}$.
  The side walls are coated with a Ni/Ti supermirror of $m=3.5$.

  The second aim is more demanding. The sample size in the scattering
  plane is given by the projection of the sample length normal to the incoming beam.
  This results in $<0.7\,\textnormal{mm}$ ($\alpha_i^s < 4^\circ$, sample
  length 10\,mm). So pre-image and sample position define the
  foci of the elliptic shape of the reflector. The spacial constraints
  on Amor lead to a focus-to-focus distance of approximately 4\,000\,mm. 
  The half axes parameters are $a_{xz} = 2000\,\textnormal{mm}$ and
  $b_{xz} = 50\,\textnormal{mm}$.
  The actual length of the device is given by the tolerable reflection angle
  for short wavelengths and the divergence to be collected. 
  The \textsl{short wavelength} limit was defined to be 4\,\AA,
  which is at the flux maximum of the wavelength distribution 
  behind the straight guide. And the maximum divergence delivered by
  the guide to a position behind the chopper shielding 
  is $\Delta\alpha_i \approx 1.6^\circ$.
  Simulations and analytical calculations both showed an optimum 
  length of 2\,000\,mm, where the distances from the guide ends to both foci
  are 1\,000\,mm. The complete length is coated with a Ni/Ti supermirror
  of $m=5$. 
  A sketch of the guide geometry is show in figure \ref{f_guide}.

   \begin{figure}[t]
     \center{\includegraphics[width=75mm]{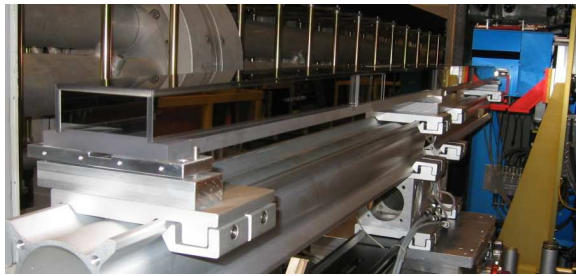}}
    \caption{\label{f_support}
     The guide mounted on the support, with screws
     to adjust the distance guide to platforms. 
    }
   \end{figure}

  The design of the elliptic guide element was done analytically and
  refined by McStas simulations. The manufacturing of the glasses,
  the $m=5$ coating and the assembly was done by SwissNeutronics;
  The $m=3.5$ coating was performed at PSI.

  %. . . . . . . . . . . . . . . . . . . . . . . . . . . . . . . . . . . . . . . 
  
   A knife edge diaphragm is installed in the centre of the elliptic
   guide to prevent the direct view in the large divergence setting and
   to allow for small divergences in cases where off-specular measurements
   or a clear resolution $\Delta q_z$ is needed.
   The blade consists preliminary of a 1\,mm BorAl sheet, followed by a 
   1\,mm Cd sheet. It is slightly wider than the guide and it is lead in
   grooves in the side walls. 
   The diaphragm is motorised and can be varied in the range
   0.1\,mm to 30\,mm, with an accuracy of 0.02\,mm.

 %..............................................................................
 \subsection{sample}

   For these tests the elliptic guide occupies the
   sample table. Thus the sample is mounted on the analyser stage.
   This allows for tilting ($\theta^s$) and lifting ($z$ direction)
   the sample and thus is sufficient for non-polarised measurements.
   
   For the final set-up intended for Amor the sample will be mounted 
   as in the normal mode on the sample table, allowing for heavy
   equipment like the 1\,T electromagnet, the horizontal 5\,T 
   cryomagnet, or the Langmuir trough.

   The results shown below where obtained using a Ni film of 1000\,\AA\ 
   thickness on glass; and a $m=5$ Ni/Ti supermirror, also on glass.
   In addition a perowskite type multilayer and a bi-block-polymer film
   were used in the monochromatic mode.

 %..............................................................................
 \subsection{detector}

   Amor is equipped with 2 $^3$He single detectors (diameter 10\,mm, 
   active length 100\,mm) and a $^3$He wire detector with an active
   window size of $180 \times 180\,\mathrm{mm}^2$. The latter has a 
   spacial resolution of $\approx 2\,\textnormal{mm}$.
 
%-------------------------------------------------------------------------------
\section{experiments}

 A first series of tests of the selene set-up was performed on Amor
 2.\,-6.\ December 2010.
 
  The set-up and pre-alignment was realised with white light
  coming from a slight projector and coupled into
  the neutron beam-path via a Si wafer located between the chopper housing 
  and the $\mathrm{1^{st}}$ slit. 
  The light covers the complete height, but only the inner 
  part in the horizontal plane. 

    \begin{figure}[b!]
     \center{%
     \includegraphics[width=60mm]{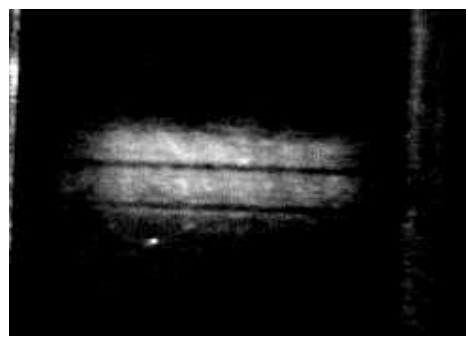}%
     }
     \caption{\label{f_stripes}
       Stripe pattern on the surface of the detector, obtained with
       white light emerging from the slit at the first focal point.
       The dark stripes can be related to the glass-glass junction area  
       in the guide.
     }
    \end{figure}

    \begin{figure*}[b!] 
     % amor_selene_mono.tex

   %\begin{figure*}[!t]
   \begin{picture}(160,40)
     \psset{unit=1mm, xunit=0.015mm, yunit=.15mm}
     \put(110,21){
       % slits
       \psline(-5242,-22.5)(-5242,37.5)% slit 1
       \psline(-3540,-30.0)(-3540,30.0)% slit 2
       \psline(-3080,-30.0)(-3080,30.0)% slit 3
       \psline( -500,-90)( -500,-30)% slit 4
       \psline( 1600,-40)( 1600,40)% slit 5
       % boxes & shadows
       \psset{linestyle=none,fillstyle=solid}%
       \psline[fillcolor=gray](-6669,-80)(-6669,80)(-5435,80)(-5435,-80)%
       \psline[fillcolor=yellow](-7200,32.5)(-6744,32.5)(-4080,7.5)(-6744,-17.5)(-7200,-17.5)%
       \psline[fillcolor=yellow](-4000,0)(-2960,13)(-2500,9)(-2000,1)(-1500,-9)(-1040,-23)%
       \psline[fillcolor=yellow](-2960,13)(-1040,-23)(0,-80)(2000,40)(2000,0)(0,-80)%
       % absorber
       \psset{linestyle=solid,fillstyle=none,linecolor=black,linewidth=0.4}%
       \psline(-4130,11.0)(-4130,31.0)%
       \psline(-3970,-2.7)(-3970,-22.7)%
       \psline(-4000,1)(-4000,21)% knife mono
       \psline(-2000,-16)(-2000,-40)% knife selene
       \psline[linewidth=1](-5500,27)(-5500,58)% lead
       \psline[linewidth=1](-5500,-12)(-5500,-43)%
       % the grid
       \psset{linestyle=solid,fillstyle=none,linecolor=blue,linewidth=0.01}%
       \psline(-7200,7.5)(-4080,7.5)(-4000,0)(-2000,0)(0,-80)(2000,20)%
       \psline{->}(-7200,-120)(2200,-120)%
        \uput[0](2200,-120){\small\color{blue}$x / \textnormal{mm}$}%
       \psline(-6744,-130)(-6744, -30)%
        \uput[90](-6744,-165){\small\color{blue}-6744}%
       \psline(-4000,-130)(-4000, -10)%
        \uput[90](-4000,-165){\small\color{blue}-4000}%
       \psline(-3000,-130)(-3000,-100)%
        \uput[90](-3000,-165){\small\color{blue}-3000}%
       \psline(-1000,-130)(-1000,-100)%
        \uput[90](-1000,-165){\small\color{blue}-1000}%
       \psline(    0,-130)(    0,-100)%
        \uput[90](    0,-165){\small\color{blue}0}%
       \psline{->}(  0, -60)(  0,   0)%
        \uput[90](0,0){\small\color{blue}$z$}%
       \psline( 2000,-130)( 2000,-100)%
        \uput[90]( 2000,-165){\small\color{blue}2000}%
       % guides
       \psset{linecolor=red}%
       \psline[linewidth=0.6](-7200,33.5)(-6744,33.5)%
       \psline[linewidth=0.6](-6744,-18.5)(-7200,-18.5)%
       \rotatebox{349}{\psellipticarc[linewidth=0.4,showpoints=false]{-}(-1850,-75)(2000,40){63}{125}}%
       % monochromator
       \psline[linewidth=0.4](-4130,11.0)(-4030,6.0)%
       \psline[linewidth=0.4](-4030,-0.3)(-3970,-2.7)%
       % frame overlap mirror
       \psline[linewidth=0.2](-4950,21.2)(-4450,-5)%
       % detector
       \psline[linewidth=1.0,linecolor=blue](2000,-10)(2000,45)%
       % labels
        \uput[90](-7020,  77){\small\color{red}guide}%
        \uput[90](-6052,-115){\small\color{gray}chopper}%
        \uput[90](-2000,  77){\small\color{red}focusing guide}%
        \uput[90](-4000,  80){\small\color{red}monochromator}%
        \uput[90](    0,  77){\small\color{black}sample}%
        \uput[90]( 2000,  80){\small\color{black}detector}%
     } %off-set
   \end{picture}
  %\end{figure*}
     \caption{\label{f_amor_selene_mono}
      Sketch of Amor in the scattering plane with the test set-up for
      selene in monochromatic mode. 
      The colour code is: 
      \textsl{black}: diaphragms, absorber; 
      \textsl{blue}: auxiliary lines and reference beam;
      \textsl{red}: (coated) guide;
      \textsl{yellow}: (part of) beam.
      The vertical axis is stretched by 10.
     }
    \end{figure*}
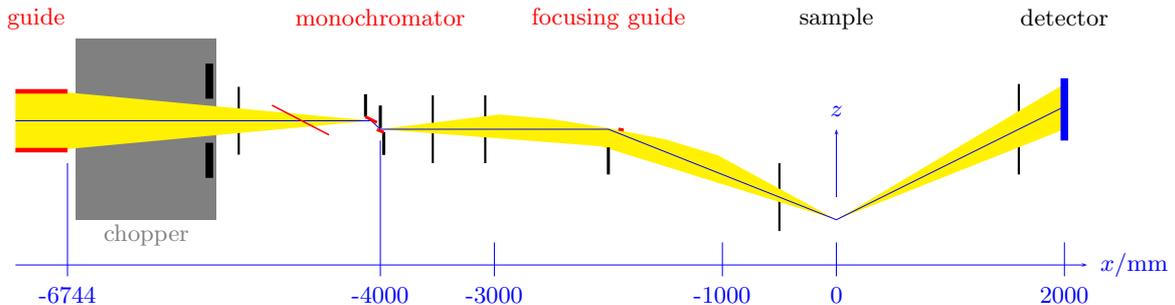

  % . . . . . . . . . . . . . . . . . . . . . . . . . . . . . . . . . . . . . . . 
  \subsection{guide: geometry and alignment}
 
   All equipment on the optical bench not needed was removed
   (i.e. the polariser, slit system 3, sample $z$-translation).
   The guide was mounted on an improvised support of X95 profiles,
   where it was fixed on its centre platform and 
   screws in the end-platforms were used to adjust it
   (see figure \ref{f_support}).
   The guide was then aligned with light collimated by slits and
   by the monochromator. The height and tilting were adjusted to 0 by feeding
   the beam through openings below the guide, so that the beam and the bottom
   of the guide were exactly parallel. Then the guide was lowered by 50\,mm
   so that its centre on the upper surface
   is at the same height as the centre of the monochromator.
   Afterwards it was tilted by $\theta^e = 1.43^\circ$ to bring the 
   $\mathrm{1^{st}}$ focal point to the position
   of the initial aperture.
   In addition the guide was aligned vertically using a narrow light beam.

   With the elements located at their nominal positions according to
   figure \ref{f_amor_selene_mono}, it was possible to track the light beam to a 
   vertical focal spot some 200\,mm upstream from the nominal position. 
   In addition the divergent beam behind the guide showed horizontal
   stripes which could be associated with the joints of the four 500\,mm
   long individual glass substrates. These stripes can be seen on the
   detector front in figure \ref{f_stripes}.
   The vertical focusing seemed correct with a spot 4000\,mm behind the 
   monochromator.

   From several tests to improve the guide geometry and form the measurements
   discussed below it looks like the guide has the following properties:
   \setlength{\leftmargini}{\parindent}
   \begin{itemize}
    \item The inner parts (along $x$) of all segments have
     the correct geometry, both in horizontal and vertical direction.
     There the waviness and roughness is sufficiently small so it does not
     affect the working principle.
    \item In horizontal direction, the 4 segments have a common
     focal spot, 1010\,mm behind the end of the guide. This is sufficiently
     close to the specified 1000\,mm since in horizontal direction the 
     initial aperture is of the order of 20 to 40\,mm.
     Also here one can see dark stripes, but these do not affect the 
     performance of the guide.
    \item In vertical direction the four segments could not be properly
     aligned. Segments 1 and 2 can be tuned
     to share a focal area some 160\,mm upstream the nominal position.
     Dark stripes are still visible since the surface does not follow the
     elliptic shape close to the ends of the segments.
     Segments 3 and 4, which produce very sharp spots in their
     focal distances, are tilted against each other and against the
     first two. The consequence is that the sample could not be adjusted
     in a way to ideally be illuminated by all 4 segments.
     This is the reason for the
     lower intensity of the reflected beam for higher angles (see figure
     \ref{f_selene_tof_map}).
    \item The junction right in the centre of the guide causes the 
     most severe problems. There the surface of the guide seems to be 
     \textsl{S}-shaped with the consequence, that the neutrons are not 
     just reflected off the sample position. That would just reduce the
     statistics. But also neutrons from a wrong direction
     reach the sample, which spoils the $\lambda(\alpha_i)$ encoding. 
     This can be seen \textsl{nicely} in figure \ref{f_selene_ml_ni}.

     In addition the dark region in the centre prevents the operation 
     mode with a small aperture. This would be the fall-back
     option to the conventional operation, allowing for alignment, and
     --- more severe --- for off-specular measurements.
   \end{itemize} 

   The consequences are, that (I) for these tests the horizontal focusing could
   not be used; (II) the data collected in the monochromatic mode could not be
   reduced; and (III) the reference measurement with the guide but a small
   inner aperture could not be performed.
   Nevertheless, the principles could be checked and in the case of the 
   TOF mode very good results were obtained, both qualitatively and 
   quantitatively.

  % - - - - - - - - - - - - - - - - - - - - - - - - - - - - - - - - - - - - - -
  \subsection{monochromatic mode}

   The double ML monochromator was mounted
   on the polariser stage and adjusted with $\theta^m = 0^\circ$ in a way that
   a collimated beam (defined by the initial slits) just passes through the opening.
   The knife edge diaphragm is 0.75\,mm from the $\mathrm{2^{nd}}$ MLs surface,
   defining a virtual aperture of 1.5\,mm height.
   Then the monochromator was tilted and shifted to the operation position
   $\theta^m = 3^\circ$, $z = -7.2\,\textnormal{mm}$.
   Accordingly all subsequent components had to be reposition 
   to $z = -7.2\,\textnormal{mm}$.
   A sketch of the set-up is shown in figure \ref{f_amor_selene_mono}.

   The measurements in the monochromatic mode were all done with the chopper
   running at 700 rpm. Since the slit reducing the beam
   (the one in the monochromator)
   was outside of all shielding, the radiation level around the instrument
   would have been too high other ways.

    \begin{figure*}[t]
     \parbox[b]{80mm}{
     \begin{picture}(80,82)
       \put(0,34){\includegraphics[width=84mm]{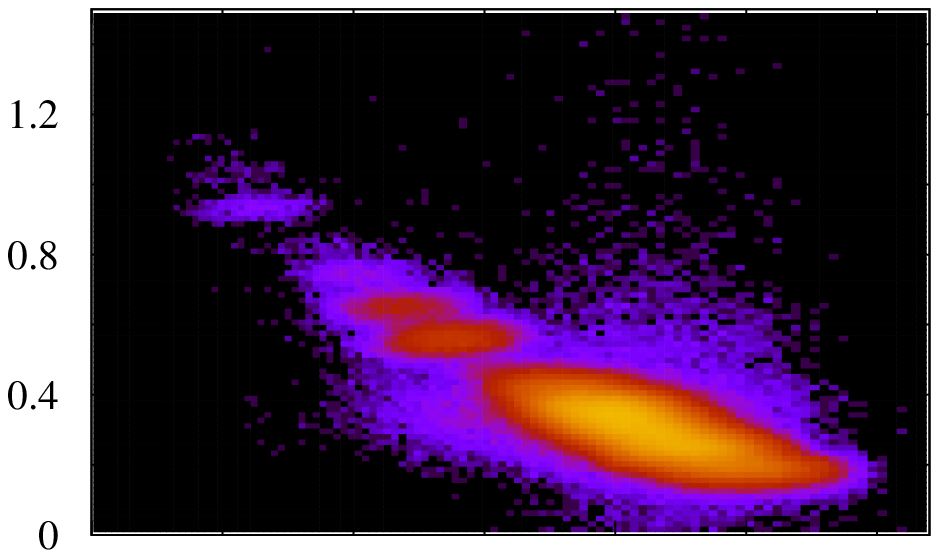}}
       \put(0,-3){\includegraphics[width=84mm]{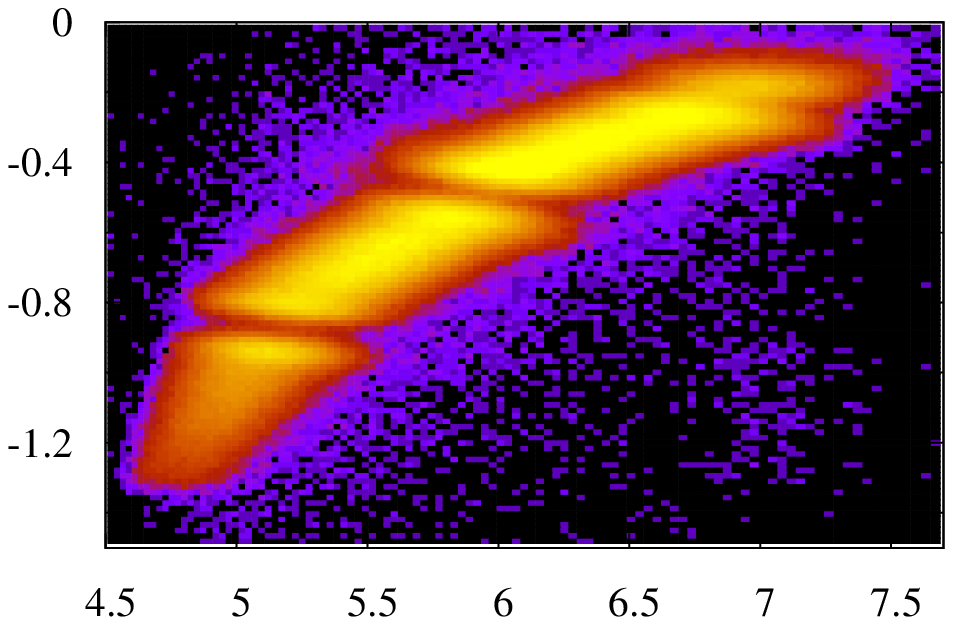}}
       \put(36,0){\small$\lambda \,/\,\textnormal{\AA}$}
       \put(0,62){\small$\alpha_f^\textnormal{s}$}
       \put(0,26){\small$\alpha_f^\textnormal{s}$}
     \end{picture}
     \caption{\label{f_selene_ml_map}
       Intensity maps over $\lambda$ and $\alpha^\textnormal{s}_f$ for the
       1000\,\AA\ Ni film on glass with 
       $\theta^s=0.5^\circ$ (\textsl{top}),
       and measured directly by replacing a slit for the sample
       (\textsl{bottom}).
     }
     }\hfill\parbox[b]{80mm}{
    %\end{figure}
    %\begin{figure}[b]
     \input{f_selene_ml_ni}
     \caption{\label{f_selene_ml_ni}
      Reflectivity of a 1000\,\AA\ Ni film on glass measured in the
      conventional TOF mode (\textsl{red}) and with the selene
      set-up in monochromator mode for various $\theta^\textnormal{s}$. 
      For the latter the normalisation 
      was not properly performed (the reference measurements are missing)
      which explains the deviations for larger $q_z$.
      But one clearly sees the gaps (marked by asterisks) 
      and pile-ups (marked with arrows)
      due to the optical errors of the elliptic guide.
     }
     }
    \end{figure*}

    \begin{figure*}[b!]
     % amor_selene_mono.tex
\begin{picture}(160,40)
 \psset{unit=1mm, xunit=0.015mm, yunit=.15mm}
 \put(110,21){
   % slits
   \psline(-4000,-22.5)(-4000,37.5)% slit 1
   \psline( -500,-82.5)( -500,-22.5)% slit 4
   \psline( 1600,-32.5)( 1600,47.5)% slit 5
   % boxes & shadows
   \psset{linestyle=none,fillstyle=solid}%
   \psline[fillcolor=gray](-5427,-80)(-5427,80)(-4193,80)(-4193,-80)%
   \psline[fillcolor=yellow](-5958,32.5)(-5502,32.5)(-4000,7.5)(-5502,-17.5)(-5958,-17.5)%
   \psline[fillcolor=yellow](-4000,7.5)(-2960,20.5)(-2500,16.5)(-2000,8.5)(-1500,-1.5)(-1040,-15.5)%
   \psline[fillcolor=yellow](-2960,20.5)(-1040,-15.5)(0,-72.5)(2000,47.5)(2000,7.5)(0,-72.5)%
   % absorber
   \psset{linestyle=solid,fillstyle=none,linecolor=black,linewidth=0.4}%
   \psline(-2000,-8.5)(-2000,-32.5)% knife selene
   \psline[linewidth=1](-4258,27)(-4258,58)% lead
   \psline[linewidth=1](-4258,-12)(-4258,-43)%
   \psline[linewidth=0.3,linestyle=dashed](-4670,-50)(-4670,65)% chopper disc 1
   \psline[linewidth=0.3,linestyle=dashed](-5160,-50)(-5160,65)% chopper disc 1
   % the grid
   \psset{linestyle=solid,fillstyle=none,linecolor=blue,linewidth=0.01}%
   \psline(-5958,7.5)(-2000,7.5)(0,-72.5)(2000,27.5)%
   \psline{-}(-5958,-120)(960,-120)%
   \psline{->}(1040,-120)(2200,-120)%
   \uput[0](2200,-120){\small\color{blue}$x / \textnormal{mm}$}%
   \psline(-5502,-130)(-5502, -30)%
   \uput[90](-5502,-165){\small\color{blue}-5502}%
   \psline(-4880,-130)(-4880,-100)%
   \uput[90](-4880,-165){\small\color{blue}-4880}%
   \psline(-4000,-130)(-4000, -30)%
   \uput[90](-4000,-165){\small\color{blue}-4000}%
   \psline(-3000,-130)(-3000,-100)%
   \uput[90](-3000,-165){\small\color{blue}-3000}%
   \psline(-1000,-130)(-1000,-100)%
   \uput[90](-1000,-165){\small\color{blue}-1000}%
   \psline(    0,-130)(    0,-100)%
   \uput[90](    0,-165){\small\color{blue}0}%
   \psline{->}(  0, -60)(  0,   0)%
   \uput[90](0,0){\small\color{blue}$z$}%
   \psline( 920,-130)( 1000,-110)%
   \psline( 1000,-130)( 1080,-110)%
   \psline( 2000,-130)( 2000,-110)%
   \uput[90]( 2000,-165){\small\color{blue}3783}%
   % guides
   \psset{linecolor=red}%
   \psline[linewidth=0.6](-5958,33.5)(-5502,33.5)%
   \psline[linewidth=0.6](-5502,-18.5)(-5958,-18.5)%
   \rotatebox{349}{\psellipticarc[linewidth=0.4,showpoints=false]{-}(-1850,-67.5)(2000,40){63}{125}}%
   % frame overlap mirror
   \psline[linewidth=0.2](-3708,21.2)(-3208,-5)%
   % detector
   \psline[linewidth=1.0,linecolor=blue](2000,-2.5)(2000,52.5)%
   % labels
   \uput[90](-5778,  77){\small\color{red}guide}%
   \uput[90](-4810,-115){\small\color{gray}chopper}%
   \uput[90](-2000,  77){\small\color{red}focusing guide}%
   \uput[90](-4000,  80){\small\color{black}slit}%
   \uput[90](    0,  77){\small\color{black}sample}%
   \uput[90]( 2000,  80){\small\color{black}detector}%
 } %off-set
\end{picture}
     \parbox{160mm}{
     \caption{\label{f_amor_selene_tof}
      Sketch of Amor in the scattering plane with the test set-up for
      selene in TOF mode.
      The colour code is:
      \textsl{black}: diaphragms, absorber;
      \textsl{blue}: auxiliary lines and reference beam;
      \textsl{red}: (coated) guide;
      \textsl{yellow}: (part of) beam.
      The vertical axis is stretched by 10.
     }
     }
    \end{figure*}
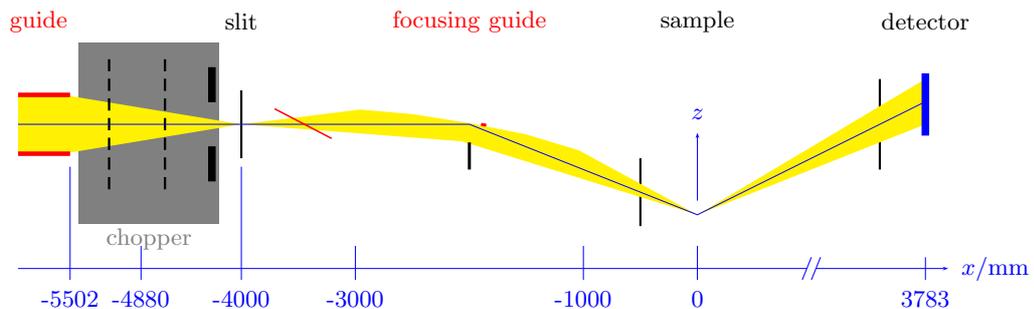

   Figure \ref{f_selene_ml_map} shows intensity maps obtained from the
   beam going directly to the detector (no sample), and reflected off of a 1000\,\AA\
   Ni film on glass, respectively.
   The black horizontal stripes in the direct measurements
   are caused by the junction areas of the elliptic guide not following
   the nominal curve. Most part of the missing intensity would pass the
   sample and thus just leads to a reduced statistics. But at least some part
   of the wrongly scattered intensity makes its way to the sample and
   thus spoils the $\lambda(\alpha)$ encoding as can be seen in figure
   \ref{f_selene_ml_ni}. 
   The additional stripe pattern for the Ni film measurement in figure 
   \ref{f_selene_ml_map} is caused by the Kiessig oscillation of the
   reflectivity and thus the expected signal in this case.

    With this set-up several samples were measured 
    with the result that the dynamic range seems to be at least 5
    orders of magnitude. The problems with the guide's alignment prevented
    a proper and quantitative analysis of these measurements.

   % - - - - - - - - - - - - - - - - - - - - - - - - - - - - - - - - - - - - - -
   \subsection{TOF mode}

    To increase the divergence of the incoming beam to $\approx 1.6^\circ$,
    the guide was moved closer to the chopper housing 
    (see figure \ref{f_amor_selene_tof}). 
    At the $1^{st}$ focal point a diaphragm is positioned to define the 
    pre-image height of 1\,mm. 
    Even with this rather small aperture one had to take care not to oversaturate
    the area detector, some 8\,m away.

\begin{figure}[t]
     \input{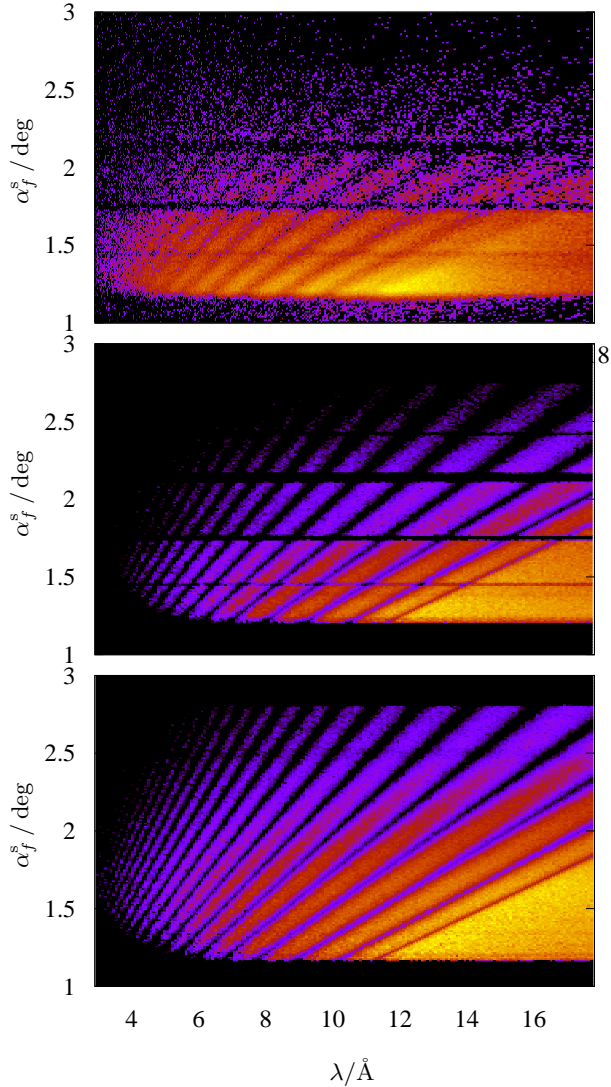}
      \caption{\label{f_selene_tof_map}
        Intensity maps 
        over $\lambda$ and $\alpha^\textnormal{s}_f$ for 
        a 1000\,\AA\ Ni film on glass, obtained with 
        a mean angle of incidence $\theta^s=1.75^\circ$.
        The intensity is given on $\log_{10}$ scale,
        where the colour spectrum runs from black ($\log_{10}I=-3$) to
        yellow ($\log_{10}I=0$).
        The \textsl{upper} map is as measured and compares to
        the simulated one in the \textsl{middle}. 
        To reproduce the measured intensity distribution in the simulation,
        the guide segments had to be tilted up to $0.03^\circ$ relative
        to each other, shifted in the $\mu$m range, and gaps of
        several cm length were assumed.
        The \textsl{lower} map is also simulated, assuming
        a perfect alignment and no gaps.
        The extra horizontal stripes in simulation and measurement
        (e.g. at $\alpha_f^\textnormal{s} \approx 1.3^circ$ and $2.5^\circ$
        are caused by features of the straight guide.
      }
    \end{figure}

    Due to the limited time, just the Ni film and the $m=5$ SM were measured,
    and only at $\theta^s = 1.25^\circ$ and $\theta^s = 1.75^\circ$.
    Figure \ref{f_selene_tof_map} shows one intensity map and the 
    corresponding result of a McStas simulation.
    The comparison of both shows the effect the miss-alignment of
    the real guide: The alignment was optimised for the 
    first segment of the guide, leading to lowest $\alpha^\textnormal{s}_i$.
    There a good agreement is found with the simulation.
    The relative tilt of the other segments leads to a shift of the focal point
    and thus to reduced intensity and to angular errors.

\begin{figure}[t]
     \input{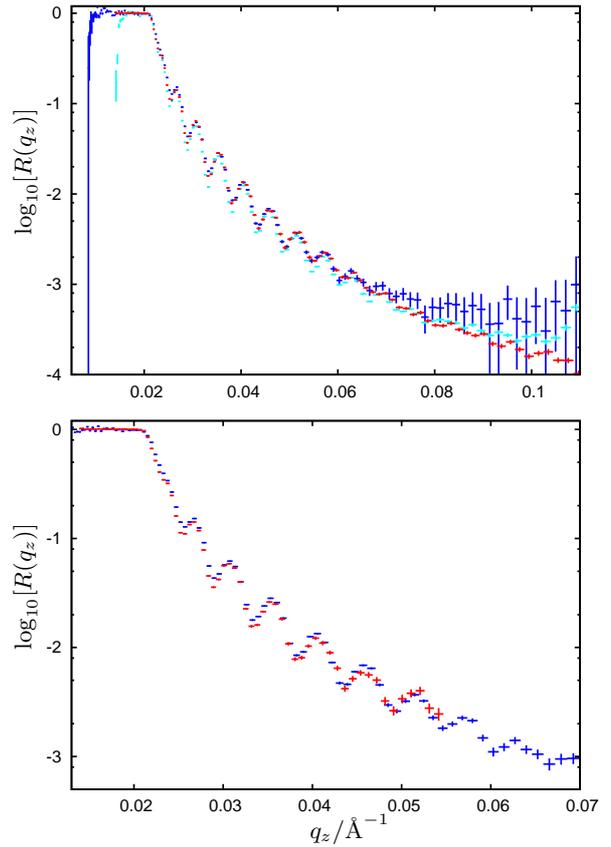}
     \caption{\label{f_selene_tof}
       \textsl{Top:} $R(q_z)$ for a 1000\,\AA\ thick Ni film on glass,
       obtained from the maps shown in figure \ref{f_selene_tof_map}.
       The \textsl{red} curve is the reference measured in the conventional
       TOF mode, the \textsl{dark blue} and \textsl{light blue} curves
       correspond to $\theta^s = -1.25^\circ$ and $\theta^s = -1.75^\circ$,
       respectively.
       \textsl{Bottom:} The same $R(q_z)$ curve for $\theta^s = -1.25^\circ$,
       (\textsl{dark blue}), now compared with a single measurement in 
       conventional mode (\textsl{red}).  
     }
   \end{figure}

    For the analysis,
    the $\lambda$ / $\alpha^\textnormal{s}_f$ map was converted 
    to a $q_z$ / $\alpha^\textnormal{s}_f$ map pixel by
    pixel. I.e. every line with constant $\alpha^\textnormal{s}_f$ 
    was transformed into a $R(q_z)$ curve.
    The SM measurements were used for normalising.
    Subsequently all data points were filled in a new $q_z$ grid with
    $\Delta q_z / q_z = 0.02$. 
    The resulting $R(q_z)$ curves are shown in figure \ref{f_selene_tof} 
    together with a curve of the same sample, measured in the conventional
    TOF mode on Amor (with 2 angular settings $\alpha = 0.5^\circ$ and
    $\alpha = 1.0^\circ$). 

   The upturn of $R(q_z)$ for large $q_z$ for the selene set-up most
   likely originate from the fact that the frame-overlap mirror was not used.
   The reason is that for the selene set-up that one would have to be curved,
   which is not yet realised. 
   To reduce the influence of frame overlap, the chopper speed was
   reduced to 700\,rpm, i.e. the wavelength range was extended to 26\,\AA.
 
   The total gain factor for the same resolution is not that easy to
   extract, since only half the guide worked as expected. 
   The lower graph of figure \ref{f_selene_tof} compares a single conventional
   measurement (where the parameters on Amor are optimised for) with one 
   using selene. The difference in counting time is a factor 6.7. 
   Taking into account that the chopper speed was reduced from 1500\,rpm to
   700\,rpm (intensity scaled down by 50\%) and that the horizontal
   focusing was missing (another factor 50\%) in the second case only, one can
   estimate a gain factor of $\approx 25$ for the same resolution, and at least
   the same statistics. 
   For higher $q_z$ the gain will be less, since there the incident beam is
   broader also in the conventional set-up.
  
%-------------------------------------------------------------------------------
\section{conclusion}

 \subsection*{guide}
  The guide is the weak point of the selene concept. 
  We found that the nominal geometry in the scattering plane was chosen
  correctly, but that the guide does not fulfil the requirements.
  At least not for the monochromatic mode and for the high-resolution
  / off-specular mode with the small aperture.

  The 4 individual segments the complete guide is made of have geometrical
  deviations at the ends, and they are misaligned relative to each other.
  The consequences are a spoiled $\lambda(\alpha)$ encoding, and dark
  ranges in $\alpha$.
  Away from the ends the shape and surface quality of the guide was
  fine. So it is most likely a problem of the assembly and adjustment
  of the individual glass substrates one has to improve. 

  Normal to the scattering plane the guide works fine.

 \subsection*{monochromatic mode}
  Due to the dark area and the distortions caused by the guide it was not
  possible to get good reflectivity curves. It was proved that
  the measurements are feasible, and that gain factors of the order 10 are
  reachable, but a complete data analysis with comparison to reference
  measurements has not been done, yet.

 \subsection*{TOF mode}
  Using TOF (for encoding the wavelength) reduced the
  influence of the shortcomings of the guide dramatically. 
  It was proved that the gain for $q_z < 0.1\,\textnormal{\AA}^{-1}$
  is at least one order of magnitude, relative to the conventional
  set-up --- with the same resolution.

 \subsection*{the future}
  Only the very last measurements have been performed with the ideal
  configuration. So one should repeat the measurements in TOF mode
  also for more demanding samples and for higher $q_z$ ranges to check
  the limits of the method. I.e. the influence of off-specular scattering,
  and the dynamic range reachable.

  In addition there are new ideas:
  \begin{itemize}
   \item
    In TOF mode the measurements were performed with $\theta^s<0$ i.e. like in
    the dispersive mode with a ML monochromator. This way the highest intensity is 
    achieved for small $\lambda$ and small $\alpha^s$. By tilting the
    sample to the other side ($\theta^s>0$) one maps the highest intensity
    on small $\lambda$ and large $\alpha^s$, i.e. on high $q_z$. At the same time
    the low resolution (small $\alpha^s$) corresponds to high $q_z$ and high
    resolution to small $q_z$.
   \item
    For the remote future a better guide 
    is desirable.
    For the TOF mode one would gain the horizontal focusing and thus
    at least a factor 2 in intensity; and the fall-back option to reduce the incoming
    divergence to conventional values.
    And the monochromatic mode would be allowed for!
   \item
    An option for a new guide is to coat its walls reflecting in the sample plane
    with polarising supermirrors. 
    This way the \textsl{direct} beam would be 
    unpolarised, and the horizontally reflected ones are polarised. With an adequate
    guide and magnetisation field geometry it should even be possible to 
    magnetise both surfaces in opposite directions, allowing for a simultaneous 
    measurement of both spin states and the unpolarised beam. 
    The 3 signals are horizontally separated on the PSD.
  \end{itemize}
   
  With the presented data we fortify our recommendation to use this concept for
  new reflectometry beamlines: It allows to start with a tiny aperture, close to 
  the cold source and thus reduces radiation and shielding problems down stream.
  By using two elliptic guides in series one can avoid aberration and it is possible
  define the beam at the joining focal point, e.g. by a polariser or a  chopper.
  With the aperture reducing the divergence one has the fall-back option to 
  the conventional TOF or angle dispersive set-up, allowing for off-specular 
  reflectometry.

  But also by using the focusing guide element as an add-on for existing 
  instruments one can benefit, as shown by the flux-gain on Amor.

%-------------------------------------------------------------------------------
 \subsection*{acknowledgments}

  This research project has been supported -- by the European Commission 
  under the 7th Framework Programme through the {\em Research Infrastructures} 
  action of the {\em Capacities} Programme, Contract No: 
  CP-CSA\_INFRA-2008-1.1.1 Number 226507-NMI3; 
  -- by the Swiss National Science Fondation through the
  {\em NCCR -- Economic stimulus package:
  Neutron optical devices for small samples}; and 
  -- by SwissNeutronics.

  For ideas and discussion we thank F.\ Ott, B.\ Cubitt, P.\ B\"oni,
  U.\ Stuhr, C.\ Schanzer, M.\ Zhernenkov, H.\ Wacklin, C.\ Niedermayer, 
  and many others.

%-------------------------------------------------------------------------------

%===============================================================================
\end{document}